\documentclass[a4paper,12pt]{article}

\usepackage{ifpdf}

\newif\ifpdf
\ifx\pdfoutput\undefined
  \pdffalse
\else
  \pdfoutput=1
  \pdftrue
\fi

\RequirePackage{xspace} %
\RequirePackage{subfigure} %
\RequirePackage[centertags]{amsmath} %
\RequirePackage{amssymb}
\RequirePackage{wrapfig} %
\RequirePackage{calc} %
\RequirePackage{ifthen}
\RequirePackage{tabularx} %
\RequirePackage{flafter} %
\RequirePackage{fancyhdr} %

\ifpdf
  \RequirePackage[pdftex]{color}%
  \RequirePackage{colortbl}%
  \RequirePackage{array}%
  \RequirePackage[pdftex]{graphicx}

  \RequirePackage[ pdftex, plainpages = false, pdfpagelabels,
                 pdfpagelayout = useoutlines,
                 bookmarks,
                 breaklinks = true,
                 linktocpage,
                 pagebackref,                      % to include page numbers in bibliography
                 colorlinks = true,
                 linkcolor = blue,
                 urlcolor  = blue,
                 citecolor = blue,
                 anchorcolor = blue,
                 hyperindex = true,
                 hyperfigures
                 ]{hyperref}

\else
  \RequirePackage{color}
  \RequirePackage{colortbl}
   \RequirePackage{array}
  \RequirePackage[dvips]{graphicx}
  \RequirePackage{hyperref}
  \usepackage{rotating}
\fi

\usepackage{makeidx} % enable indexing
\usepackage{setspace} %enable setting spacing, %e.g. singlespace, onehalfspace, and doublespace
\usepackage{rotating} %enable sidewaysfigure
\usepackage{ecltree}
\usepackage{epic}
\usepackage{supertabular}  % this will allow the table not to break
\usepackage{color}
\usepackage{exscale}
\usepackage{fontenc}
\usepackage{ifthen}
\usepackage{latexsym}
\usepackage{makeidx}
\usepackage{syntonly}
\usepackage{inputenc}
\usepackage{graphicx}
\usepackage{setspace}
\usepackage{caption2}
\usepackage[english]{babel}
\usepackage[square, comma,numbers,sort&compress]{natbib}
\usepackage{hypernat}
\usepackage{boxedminipage}
\usepackage{framed}
\usepackage{longtable}
\usepackage[all]{hypcap}    %included to make the hyperlink go to the top of figure
\usepackage{algorithm2e}
\usepackage{algorithmic}
\usepackage{lscape}
\usepackage{pdflscape}
\newcommand{\note}[1]{\marginpar[left]{\singlespace \tiny #1}}

\renewcommand{\sectionmark}[1]%
      {\markright{\thesection\ #1}} %stops it capitalizing. #1 has value of section name

\renewcommand{\note}[1]{}

\doublespace

\begin{document}
\begin{center}
{\Large Energy minimization for the flow in ducts and networks}
\par\end{center}{\Large \par}

\begin{center}
Taha Sochi
\par\end{center}

\begin{center}
{\scriptsize University College London, Department of Physics \& Astronomy, Gower Street, London,
WC1E 6BT \\ Email: t.sochi@ucl.ac.uk.}
\par\end{center}

\begin{abstract}
\noindent The present paper is an attempt to demonstrate how the energy minimization principle may
be considered as a governing rule for the physical equilibrium that determines the flow fields in
tubes and networks. We previously investigated this issue using a numerical stochastic method,
specifically simulated annealing, where we demonstrated the problem by some illuminating examples
and concluded that energy minimization principle can be a valid hypothesis. The investigation in
this paper is more general as it is based to a certain extent on an analytical approach.

\vspace{0.3cm}


\par\end{abstract}

\begin{center}

\par\end{center}

%XXXXXXXXXXXXXXXXXXXXXXXXXXXXXXXXXXXXXXXXXXXXXXXXXXXXXXXXXXXXXXXXX
\section{Introduction} \label{Introduction}

One of the general principles that governs the behavior of many physical systems is energy
minimization where the system tends to reduce its energy to the lowest available state within the
given constraints. Examples of the occurrence of this principle are the inclination of a massive
body permeated by an external gravitational field to take the position of the lowest available
potential energy, and the tendency of atomic and molecular multi-level systems to decay to their
lowest allowed energy level which is usually the ground or a meta-stable state. Therefore, it is
logical to presume that such a principle may also govern the behavior of flow systems like fluid
transport devices and electric power distribution networks.

In a previous investigation \cite{SochiPresSA2014}, it was hypothesized that the dynamical
attributes of the flow systems are subject to the energy minimization principle. This assumption
was demonstrated by resolving the flow fields correctly in a number of typical fluid systems using
the minimization principle in conjunction with a simulated annealing \cite{MetropolisRRTT1953,
KirkpatrickGV1983, Cerny1985} numerical protocol. Some criticism that may be directed to the
previous attempt is the lack of mathematical rigor since the validity of the energy minimization
principle was only demonstrated by a few examples representing single tubes and networks of
interconnected tubes. The stochastic nature of the simulated annealing approach may also be
criticized due to possible contamination of the obtained solutions in some cases with considerable
numerical errors that may cast doubt on the reliability and validity of the method and the
underlying principle.

Although we do not believe that these criticisms undermine the main objective of the previous
investigation which is largely based on demonstrating the viability of the proposed hypothesis
while relying for the final conclusion for the validity of the energy minimization principle on the
vanishing likelihood of obtaining these correct solutions by a providential coincidence, we believe
that a more rigorous approach is needed to establish the principle. Hence a more rigorous treatment
of analytical nature is attempted in the current paper to establish the energy minimization
principle as a governing rule in these flow systems.

The current investigation will focus on the fluid dynamic systems represented by single ducts and
networks of interconnected ducts which are widely used in fluid transportation. However, the energy
minimization argument is more general as it can be extended to any flow system whose physical
behavior can be described by a mathematically equivalent formulation to the one that applies to the
examined fluid systems such as conducting electronic components and power distribution networks.
The current investigation is also based on the assumption of linear flow elements which encompass
rigid conduits of uniform cross sections, characterized by constant cross sectional shape and area
along their axial direction regardless of any imposed pressure. This condition, besides the
forthcoming assumptions about the type of flow and fluid, will ensure the linearity of the pressure
drop as a function of the axial coordinate in the flow direction in the individual elements of
these flow systems.

Regarding the nature of the flow, we assume a laminar, incompressible, pressure-driven,
fully-developed flow with minor entry and exit edge effects and negligible viscous frictional
losses. We also assume negligible effects from external body forces like gravity and electric and
magnetic attraction or repulsion. As for the boundary conditions, we assume Dirichlet-type pressure
boundary conditions; although the energy minimization argument, if established, will not be
restricted to such conditions which are mainly based on the way that the flow system is formulated
and described by the observer and hence the type of these conditions does not represent an
intrinsic property of the system that is due to determine its ultimate outcome. Concerning the type
of fluid, we assume a generalized Newtonian fluid with a power law dependency of its volumetric
flow rate on the pressure drop in a pressure-driven flow which include, among other possibilities,
Newtonian and Ostwald--de Waele fluids.

For clarity, we also state the obvious assumption that the pressure dependency of the volumetric
flow rate in any conducting element is restricted to the inlet and outlet pressures of that element
and not on the pressure at other points in the network although the pressure values in the network
as a whole are generally correlated to each other as they are all subject to a single equilibrium
constraint that requires the adjustment of the pressure field on a global scale to ensure a stable
configuration.

The purpose of most of the above-stated restrictions regarding the flow, fluid and conduit is to
facilitate the energy minimization argument. However, some of these restrictions may be relaxed
although we will not try to do so in the present paper. Further investigations in the future may
examine some of these extensions and generalizations.

%XXXXXXXXXXXXXXXXXXXXXXXXXXXXXXXXXXXXXXXXXXXXXXXXXXXXXXXXXXXXXXXXX
\section{Energy Optimization Argument}

In this section we outline our argument for the support of the energy optimization as a potential
governing rule for determining the physical equilibrium state and hence the stable configuration of
the flow fields in the flow devices which include single ducts and networks of interconnected
ducts. In the next section we will discuss the nature of the energy extremum and if it is a maximum
or a minimum. The possibility of a saddle point can be ruled out by the fact that the flow fields
in these systems are stable with respect to a perturbing change and hence the stationary point that
represents the extremum cannot be a saddle point. An obvious consequence, with a practical value,
of establishing the energy optimization is the possibility of exploiting this principle to design
analytical and numerical algorithms for resolving the flow fields and obtaining the pressure and
volumetric flow rate at any point throughout the flow system.

In the previous investigation \cite{SochiPresSA2014}, we presented an argument to establish the
existence and uniqueness of the flow fields solution in conduits and networks. The argument also
implies a practical method for resolving the flow fields analytically and numerically. The essence
of that argument is that; considering the previously stated assumptions, to describe the flow in a
multi-element network, which includes a discretized single conduit as it can be regarded as a
serially-connected one-dimensional network, we set a system of $n$ simultaneous equations in $n$
unknowns where $n$ is the total number of the nodes in the network which include the boundary as
well as the internal nodes. The equations of the inlet and outlet boundary nodes are based on the
presumed boundary conditions while the equations of the internal nodes are derived from the mass
conservation principle in conjunction with the characteristic flow relation that correlates the
driving and induced fields, like pressure and volumetric flow rate in the Hagen-Poiseuille law. The
characteristic flow relation is basically derived from the momentum conservation principle in
cooperation with the fluid-structure constitutive relation.

For the linear case, in which the volumetric flow rate is linearly dependent on the pressure drop,
these equations are linearly independent, due to the fact that they cannot be represented as scalar
multiples of each other. Therefore, we have a system of $n$ independent linear equations in $n$
unknowns and hence a solution does exist and it is unique according to the rules of Linear Algebra.
For the non-linear case, the system consists of $n$ independent equations that correctly depicts a
real physical system within the given assumptions and constraints and therefore it should have a
unique solution that can be obtained iteratively by a Newton-Raphson procedure or other non-linear
solution methods. Further details about these issues are given in \cite{SochiPresSA2014}.

The new argument is based on the fact that the time rate of energy consumption, $I$, of fluid
transportation through a single conducting device, considering the type of flow systems that meet
our pre-stated assumptions, is given by

\begin{equation}
I=\Delta p\, Q
\end{equation}
where $\Delta p$ is the pressure drop across the conducting device and $Q$ is the volumetric flow
rate of the transported fluid through the device. For a flow conducting device that consists of or
discretized into $m$ conducting elements indexed by $l$, the total energy consumption rate,
$I_{t}$, is given by

\begin{equation}\label{ItEq}
I_{t}(p_{1},\ldots,p_{n})=\sum_{l=1}^{m}\Delta p_{l}Q_{l}
\end{equation}
where $n$ is the number of the boundary and internal nodes. For a single duct, the conducting
elements are the discretized sections, while for a network they represent the conducting ducts as
well as their discretized sections if discretization is employed. In Equation \ref{ItEq}, it is
assumed that $Q$ is a function of the nodal pressures ignoring the other mainly-fixed dependencies,
such as the pre-determined geometric factors, which are irrelevant to consider in this context.

The requirement of a discretization scheme in multiple-duct networks, similar to the one used for
single conduits, may arise because of the necessity of identifying the axial pressure values at the
intermediate points of the network conduits as well as the nodal pressure values at the network
main junctions. However, for the flow systems with a linear pressure drop dependency on the axial
coordinate of their components, which are the subject of the present investigation, only the nodal
pressures are required since the values of pressure at the in-between points can be obtained by a
plain linear interpolation scheme. The incompressibility of fluid ensures a constant volumetric
flow rate at any cross section of each conduit.

It is well known from Mathematical Analysis that a stationary point of a real-valued multi-variable
function, $f(x_{1},\ldots,x_{n})$, is characterized by a vanishing gradient, i.e.

\begin{equation}
\nabla f={\mathbf 0}
\end{equation}
For a real-valued $n$-variable function like $f$, the latter equation can be resolved into a set of
$n$ simultaneous equations, that is

\begin{equation}
\frac{\partial f}{\partial x_{i}}=0\,\,\,\,\,\,\,\,\,\,\,\,\,\,\,\,\,\,\left(i=1,\ldots,n\right)
\end{equation}
If this set of simultaneous equations can be solved, the stationary point, as well as the value of
the function at that point, will be identified.

When the resolved set is linear, the system can be solved directly by several standard algebraic
methods which are thoroughly investigated in the literature of Linear Algebra, while if the set was
non-linear the system of equations is usually solved by iterative methods such as the widely used
Newton-Raphson procedure. In both cases, the solution obtained directly or iteratively is unique,
assuming that the system of equations is representing a correctly formulated physical problem.
These issues are discussed in a number of our previous papers such as \cite{SochiPresSA2014,
SochiPois1DComp2013, SochiPoreScaleElastic2013} and hence they will not be pursued further.

In the current investigation, the objective is to find the extremum point of the total rate of
energy consumption function, $I_{t}(p_{1},\ldots,p_{n})$, which takes the place of the generic
$f(x_{1},\ldots,x_{n})$ function and hence we should solve the following set of simultaneous
equations

\begin{equation}\label{MainEq}
\frac{\partial I_{t}(p_{1},\ldots,p_{n})}{\partial
p_{i}}=0\,\,\,\,\,\,\,\,\,\,\,\,\,\,\,\,\,\,\left(i=1,\ldots,n\right)
\end{equation}
where $n$ is the number of all nodes in the network which include the boundary as well as the
internal nodes. It will be shown that the system of equations, given by Equation \ref{MainEq}, that
satisfies our assumptions will ultimately produce a system of $n$ simultaneous equations that have
a unique solution which is identical to the solution that is obtained from the other method which
is not based on the energy minimization principle.

The individual equations of System \ref{MainEq} can be classified into two main types: those
representing the partial derivatives of the pressure at the inlet and outlet boundary nodes and
those representing the partial derivatives of the pressure at the internal nodes. As for the first
type, we have

\begin{equation}\label{bEq}
\frac{\partial I_{t}}{\partial p_{_{b}}}=0
\end{equation}
where $b$ refers to a boundary node. Considering the fact that for a particular problem the
boundary conditions are fixed and hence are not subject to variation because they are given
conditions, the only meaning of Equation \ref{bEq} is that $I_t$ is not subject to variation with
respect to variation of $p_{_b}$ as the latter variation is not allowed. The essence of this is
that

\begin{equation}\label{bEq2}
p_{_b}=C_{_b}
\end{equation}
where $C_{_b}$ are the given constant boundary conditions.

In fact this argument regarding the boundary nodes may not be needed because the energy
minimization is considered for the specified flow system within the stated boundary constraints and
hence these are not part of the variational process that is required to optimize the energy
consumption. However, we resorted to this rather veiled mathematical reasoning for establishing the
equations for the boundary conditions to make the energy optimization argument more rigorous and
formal; otherwise we could have restricted the nodes in System \ref{MainEq} to the internal ones
and obtained the equations for the boundary nodes directly from the stated boundary constraints of
the given flow system. Anyway, both arguments regarding the boundary nodes will lead to the same
set of boundary equations which summarize the imposed boundary conditions as given by Equation
\ref{bEq2}.

As for the second type, each equation of System \ref{MainEq} will in general have multiple terms
which involve the partial derivatives of the particular internal node, as indexed by $i$, and the
partial derivatives of all its neighboring nodes which are connected to it directly through a
single duct. All the other terms in any equation of System \ref{MainEq} which do not depend
explicitly or implicitly on the pressure of the particular internal node $p_{i}$ will vanish and
hence the equation will be reduced to the following form

\begin{equation}
\frac{\partial I_{t}}{\partial p_{i}}=\frac{\partial}{\partial
p_{i}}\left[\sum_{k}\left(p_{k}-p_{i}\right)Q_{ki}+\sum_{l}\left(p_{i}-p_{l}\right)Q_{il}\right]=0
\end{equation}
where $k$ indexes the source nodes of node $i$ (i.e. $p_{k}>p_{i}$) and $l$ indexes the sink nodes
of node $i$ (i.e. $p_{i}>p_{l}$), and $Q_{ki}$ and $Q_{il}$ are the positive flow rates in the
($k\rightarrow i$) and ($i\rightarrow l$) directions respectively. On expanding the last equation
we obtain

\begin{equation}
\sum_{k}Q_{ki}\frac{\partial}{\partial
p_{i}}\left(p_{k}-p_{i}\right)+\sum_{k}\left(p_{k}-p_{i}\right)\frac{\partial Q_{ki}}{\partial
p_{i}}+\sum_{l}Q_{il}\frac{\partial}{\partial
p_{i}}\left(p_{i}-p_{l}\right)+\sum_{l}\left(p_{i}-p_{l}\right)\frac{\partial Q_{il}}{\partial
p_{i}}=0
\end{equation}
that is

\begin{equation}
\sum_{k}-Q_{ki}+\sum_{k}\left(p_{k}-p_{i}\right)\frac{\partial Q_{ki}}{\partial
p_{i}}+\sum_{l}Q_{il}+\sum_{l}\left(p_{i}-p_{l}\right)\frac{\partial Q_{il}}{\partial
p_{i}}=0
\end{equation}

For the fluids whose volumetric flow rate has a power law dependency on the pressure drop in a
pressure-driven flow, such as Newtonian and Ostwald--de Waele, the second and fourth terms in the
last equation are simply constant multiples of the corresponding volumetric flow rate with the
correct sign, i.e. $-\alpha Q_{ki}$ for the second term and $\alpha Q_{il}$ for the fourth term
where $\alpha$ is a positive constant. The last equation then reduces to

\begin{equation}
\left(1+\alpha\right)\left[\sum_{k}-Q_{ki}+\sum_{l}Q_{il}\right]=0
\end{equation}
Since $\left(1+\alpha\right)\neq0$, the last equation becomes

\begin{equation}\label{massEq}
\sum_{k}-Q_{ki}+\sum_{l}Q_{il}=0
\end{equation}
which is no more than the mass conservation principle for incompressible flow.

Now, if we gather all the equations that have been obtained from the energy optimization argument,
which include the boundary nodes equations as well as the internal nodes equations, we obtain a
system of simultaneous equations which is identical to the system that has been obtained from the
argument based on the conservation principles in conjunction with the fluid-structure constitutive
relations. Whether the system is linear or non-linear in pressure, it has a unique consistent
solution, as outlined previously in this paper and in \cite{SochiPresSA2014}, and hence the
solution obtained from the energy optimization principle is the same as the solution obtained from
the conservation principles.

%XXXXXXXXXXXXXXXXXXXXXXXXXXXXXXXXXXXXXXXXXXXXXXXXXXXXXXXXXXXXXXXXX
\section{Type of Extremum}

We now discuss the issue related to the nature of the energy extremum point found in the last
section and if it is a minimum or a maximum, excluding the possibility of a saddle point as
indicated earlier.

There are several arguments that can be given to support the fact that the extremum is a minimum
point of the system. First the energy minimization principle is a general rule that governs many
physical systems and hence if energy optimization principle is established in a system it is more
likely to be a minimizing principle. Second, as shown in \cite{SochiPresSA2014} the same solutions
can be obtained from the simulated annealing method which is fundamentally based on a minimization
principle. Perturbation arguments may also be proposed in support of minimization where slight
changes in the equilibrium solution will produce a rise in the total energy of the system. We may
also notice that global maximum energy is not bounded and hence it is not a possibility for a
confined physical system. More elaboration on the arguments in support of the minimal energy
proposal is presented in \cite{SochiPresSA2014}. We therefore conclude that the extremum found by
the argument presented in the last section, by resolving the pressure field, is a minimum energy
point of the flow system rather than a maximum.

Anyway, for all the practical purposes, such as designing solution algorithms, the value of the
main conclusion reached, specifically energy optimization, will not be compromised even if the
nature of the energy extremum is not established with certainty although the theoretical value of
the conclusion may be diminished since an extremum statement is weaker than a minimum statement.

\clearpage
%XXXXXXXXXXXXXXXXXXXXXXXXXXXXXXXXXXXXXXXXXXXXXXXXXXXXXXXXXXXXXXXXX
\section{Conclusions} \label{Conclusions}

In this paper we examined the previously investigated \cite{SochiPresSA2014} hypothesis that the
energy minimization principle is a governing rule that determines the stable configuration of the
flow fields in the fluid dynamics systems and other similar transport devices. Unlike the previous
attempt, which is founded on a rather tentative approach based on establishing the hypothesis by
examples using a stochastic numerical method where the validity of the hypothesis is deduced from
the vanishing likelihood of obtaining correct solutions by a random coincidence, in the current
investigation we resort to a rather general analytical argument to establish this hypothesis.
Although the present investigation has several limitations regarding the type of flow, fluid and
conduit, we feel that the energy minimization principle is more general than the investigated cases
although we do not attempt to establish its generality here. Future investigations could focus on
these issues among many other potential issues that are worth to investigate.

Apart from the obvious theoretical value of the main conclusion of this study regarding the energy
minimization principle as a potential governing rule in the flow systems, it can be used to design
analytical and numerical methods for resolving the flow fields. The previous investigation
\cite{SochiPresSA2014}, which is based on the use of simulated annealing, is one such example that
utilizes a numerical stochastic method. We believe that other stochastic and deterministic methods
can also be used to exploit this principle for resolving the flow fields. These methods, amongst
other related extensions and expansions, can be the subject of further investigations in the
future.

\clearpage %\vspace{1cm}
%XXXXXXXXXXXXXXXXXXXXXXXXXXXXXXXXXXXXXXXXXXXXXXXXXXXXXXXXXXXXXXXXXXX
\phantomsection \addcontentsline{toc}{section}{Nomenclature} %
{\noindent \LARGE \bf Nomenclature} \vspace{0.5cm}

\begin{supertabular}{ll}
$\nabla$                &   grad operator \\
$C_{_b}$                &   given pressure boundary conditions \\
$f$                     &   generic real valued multi-variable function \\
$I$                     &   time rate of energy consumption for fluid transport \\
$I_t$                   &   time rate of total energy consumption for fluid transport \\
$m$                     &   number of discrete elements in the fluid conducting device \\
$n$                     &   number of nodal junctions in the fluid conducting device \\
$p$                     &   pressure \\
$\Delta p$              &   pressure drop across flow conduit \\
$Q$                     &   volumetric flow rate \\
$x_i$                   &   generic independent variable \\
\end{supertabular}

\clearpage %\vspace{0.5cm}
%XXXXXXXXXXXXXXXXXXXXXXXXXXXXXXXXXXXXXXXXXXXXXXXXXXXXXXXXXXXXXXXXXXX
\phantomsection \addcontentsline{toc}{section}{References} %
\bibliographystyle{unsrt}
%\bibliography{Bibl}

\end{document}